# Using oxides to compute with heat

Guillaume F. Nataf[1], Sebastian Volz[2], Jose Ordonez-Miranda[2], Jorge Íñiguez-González[3,4], Riccardo Rurali[5], Brahim Dkhil[6]

[1] GREMAN UMR 7347, CNRS, University of Tours, INSA Centre Val de Loire, 37000 Tours, France
[2] LIMMS, CNRS-IIS IRL 2820, The University of Tokyo, Tokyo 153-8505, Japan
[3] Materials Research and Technology Department, LIST, Esch-sur-Alzette, Luxembourg
[4] Department of Physics and Materials Science, University of Luxembourg, Belvaux, Luxembourg
[5] Institut de Ciència de Materials de Barcelona, ICMAB-CSIC, Campus UAB, 08193 Bellaterra, Spain
[6] Université Paris-Saclay, CentraleSupélec, CNRS-UMR8580, Laboratoire SPMS, Gif-sur-Yvette, France

*One of the most innovative possibilities offered by oxides is the use of heat currents for computational purposes. Towards this goal, phase-change oxides, including ferroelectrics, ferromagnets and related materials, could reproduce sources, logic units and memories used in current and future computing schemes.*

Since the early days of computing, hardware has evolved from mechanical to digital systems, where logic operations are performed by con-trolling electrons in semiconductors. However, the development of semiconductor-based chips is encountering important bottlenecks. To be disruptive, future progress is expected to be driven by different information carriers — such as spins in spintronics and orbital angular momentum in orbitronics — or different computing paradigms — such as neuromorphics. For all these alternatives, oxides already show great promise. They are made of non-toxic, naturally abundant, cheap and chemically stable elements. Some of them exhibit half-metallicity and can induce an electrically tunable exchange bias in soft magnetic materials, both essential characteristics for spintronics. Some oxide heterostructures also exhibit, at their interfaces, a strong coupling between the spin, charge and lattice degrees of freedom, leading to intriguing spin–orbit interactions. In addition, the multiple resistance states of ferroelectric and ferromagnetic oxides have been used to design neuromorphic synapses. Finally, oxides are also the ideal solid-state materials for a new type of computing based on thermal currents rather than electric currents.

**Input data: writing states with heat**

Heat computing promises to be a zero-power analogue not only of conventional computing but also of innovative post-von Neumann computing schemes. It could solve many of the bottlenecks faced by electronic computing, such as high energy consumption, overheating and lack of robustness in harsh environments (such as in space, in nuclear reactors or under cold temperatures). Indeed, heat is a pervasive form of energy and is one of the main actors in a plethora of scenarios, ranging from exothermic chemical reactions to mechanical friction. It could thus be harvested from the environment and there would be no need for energy-consuming sources, as in electronic architectures. If one is to focus on electronic applications, hot spots are ubiquitous in ultra-scaled integrated circuits, where Joule heating causes temperature rises. So far, this heat has only been considered detrimental to electronic operations and a source of inefficiency, whereas it could be used as an input in thermal computing systems. Complementarily, cooling on demand is also possible by applying a voltage.

Critically, many oxides present phase transitions (elastic, polar, magnetic, electronic) near which the materials are extremely sensitive to external fields or temperature changes. This ultra-sensitivity provides us with unique ways of adjusting their thermal properties on demand. Nanostructured thermoelectric oxides have demonstrated large temperature changes under moderate voltages. Electrocaloric effects — adiabatic temperature changes and isothermal heat transfers under voltage application — have been obtained in many oxides, with relatively large temperature changes in multilayer capacitors[1].

Magnetocaloric, elastocaloric and barocaloric effects could also be envisioned as promising ways to input heat, although their implementation at the nanoscale still presents considerable challenges.

**Control and arithmetic logic units: processing with heat**

Electron-based computing relies on logic units such as switches, diodes and transistors. Building equivalent units for heat-based computing systems requires the dynamical control of heat flows. Ferroelectric oxides can permit this control, as their thermal conductivity can be manipulated with an applied electric field. Various mechanisms may enable this manipulation, for example the scattering of phonons on domain walls[2], which are natural two-dimensional topological defects that can be 'written' and 'erased' by external electric fields. One can also take advantage of important changes in the lattice structure, volume and/or anharmonicity at phase transitions[3]. Further, mechanisms based on polarization alone — such as the reorientation of polarization driven by an external field and the anisotropy of the thermal conductivity tensors[4] — can be exploited as well. Experimental realizations of some of these effects have already been demonstrated: an electric field-controlled thermal switch based on the antiferroelectric $PbZrO_3$ that operates with high switching ratios (2.2) has already been reported[3]. This ratio is twice the ratios obtained previously with electric fields at room temperature and comparable to those obtained with magnetic fields in metallic nanowires. Further, by tailoring the density of domain walls, dynamical control of the thermal conductivity has also been reported[5].

Another possibility consists in applying a magnetic field, which leads to anisotropic heat conduction — featuring giant thermal Hall effects — that has been reported for several oxides. The underlying mechanism remains debated, with some emphasizing the role of free carriers other than phonons. Alternatively, a magnetic field can also be used to trigger a phase transition between different magnetic orderings (such as antiferromagnetic to ferromagnetic, paramagnetic to ferromagnetic), which can be accompanied by a sizeable change in the thermal conductivity.

Designing a thermal rectifier (a thermal diode) is another challenge because it requires materials with asymmetric heat conduction properties. Still, by building a heterostructure that combines oxides with different thermal conductivities, a thermal diode displaying one of the highest rectification efficiencies — which quantifies the efficiency of unidirectional heat flow — was reported[6]. Additionally, in a given temperature range, some oxides exhibit a transformation from a conductive phase to an insulating one. This metal–insulator transition, when properly engineered, can give rise to anisotropic thermal conduction driving the operation of thermal diodes and thermal transistors[7].

**Thermal memory units**

Oxides exhibiting a metal–insulator phase transition have also been used as solid-state thermal memories (such as thermal memristors or memory resistors mimicking artificial synapses) of both conductive and radiative heat currents[8]. In particular, the combination of this phase transition with the thermal hysteresis — exhibiting substantial differences in the physical properties for the heating and cooling processes — allows the development of radiative thermal memristors characterized by a Lissajous curve — describing the superposition of two phase-shifted sinusoidal wavefronts enclosed by a rectangular boundary — between the heat flux and temperature periodically modulated in time. The ratio of thermal memristance between its 'on' and 'off' states is determined by the emissivity contrast between the insulating and metallic phases of the used oxides[9]. The analogy between thermal memristors and their electrical counterparts makes them promising to develop thermal computing with phonons.

**Toward proof-of-concept devices**

There is compelling evidence that oxides can enable the development of a novel scheme of computing where information is encoded with heat rather than charge carriers. Oxides tick many of the boxes of the ideal materials for such applications, and explorative theoretical proposals are being followed by promising experimental proof-of-concept devices. Thermal switches, diodes and transistors[10] are being proposed (and realized) and can be combined to provide thermal AND, OR and NOT logic gates. As an example of such transistors, gate-temperature dependent magnetic forces have been used to toggle the source–drain thermal conductance between two states and thus control the source–drain heat flow[10]. By adding thermal inputs and memory, several designs for thermal computing can be developed.

Admittedly, several challenges remain for this research field to reach its full maturity. Although enormous progress has been made for measuring temperature changes and heat fluxes, these measurements are still difficult when it comes to thin films and nanostructures featuring large tunable thermal conductivities. Furthermore, fundamental mechanisms responsible for thermal conductivity variations in response to electric and magnetic fields remain debated. This essential understanding will be instrumental in developing practical devices for computing with heat.

Until now, research in oxides for electronics and research in their thermal properties have developed independently from each other. Heat computing with oxides should thus receive greater attention to foster mutually beneficial research. The proposal for a heat-based neuromorphic system (Fig. 1) is a striking example of the possibilities offered by such cross-fertilization. It shows how oxides could be effectively used to design synapses driven by heat only (Fig. 1a,b) or by a combination of heat and voltage (Fig. 1c,d). They could take advantage of the bulk material properties (phase transition induced by temperature) or local spontaneous nanostructuration (domain walls controlled by an applied electric field).

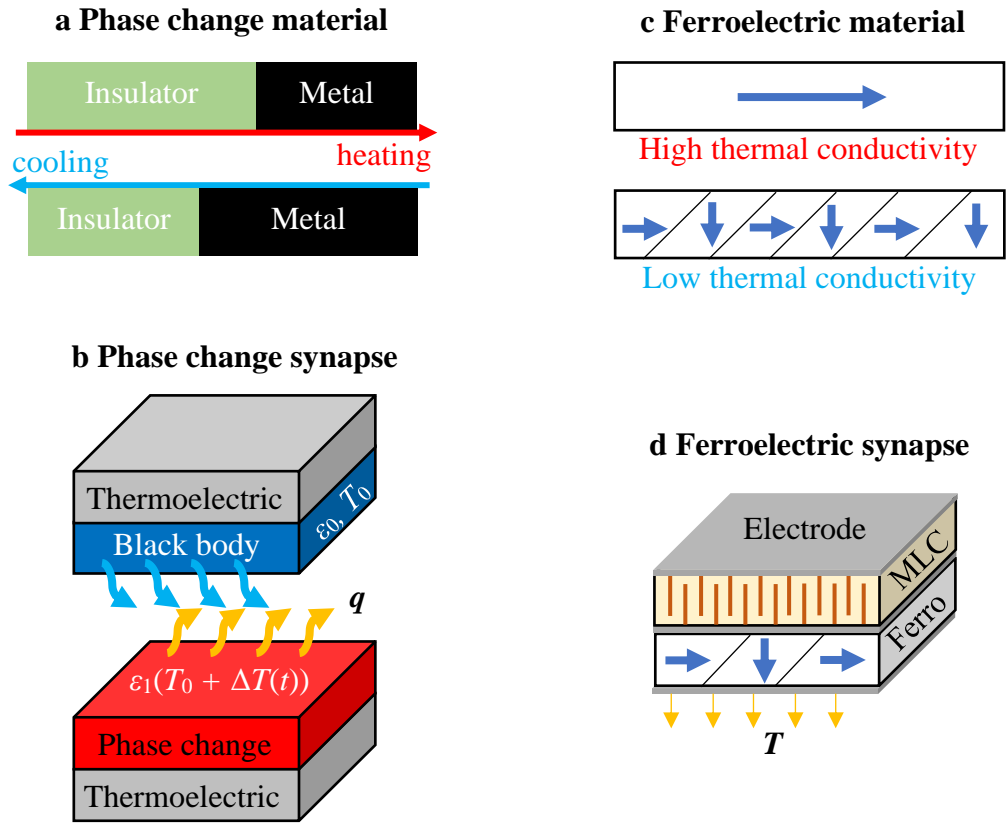

**Figure 1. Examples of heat-based neuromorphic systems. a** A phase-change material with a low-temperature insulating state and a high-temperature metallic state, exhibiting thermal hysteresis. **b** A synapse based on the time-dependent (t) temperature modulation (via a Peltier element) of a phase-change material[9], taking advantage of its emissivity ($\varepsilon$) difference during heating and cooling across its phase transition, which leads to different heat flows ($q$). **c** A ferroelectric material with different percentages of switched domains (and/or density of domain walls), leading to different thermal conduction states. **d** Mixed electronic–thermal synapse where spike inputs are provided by an electrocaloric multilayer capacitor (MLC)[1]. Trains of voltage pulses result in trains of heat and/or cold. The voltage pulses also gradually switch the polarization in the ferroelectric material, resulting in different percentages of switched domains and in finely tuned thermal conduction states[2,4], mimicking synaptic weights. The output of the neuromorphic system can be read as a temperature $T$.

**Acknowledgments**
This Comment was cofunded by the European Union [European Research Council (ERC), DYNAMHEAT, No. 101077402; European Program FP7-NMP-LARGE-7, QUANTIHEAT, No. 604668]. Views and opinions expressed are, however, those of the authors only and do not necessarily reflect those of the European Union or the ERC. J.Í.G. acknowledges the funding of the Luxembourg National Research Fund (FNR) through Grant C21/MS/15799044/FERRODYNAMICS. R.R. acknowledges financial support by MCIN/AEI/10.13039/501100011033 under grant PID2020-119777GB-I00, and the Severo Ochoa Centres of Excellence Program under grant CEX2019-000917-S, and by the Generalitat de Catalunya under grant 2021 SGR 01519.

**Competing interests**
The authors declare no competing interests.